\documentclass[aps,prd,notitlepage,longbibliography,
superscriptaddress,nofootinbib,amsmath,amssymb,11pt]{revtex4-2}

\usepackage[]{inputenc}
\usepackage{color,xcolor}
\usepackage{bm}
\usepackage{mathtools}
\usepackage{physics}
\definecolor{linkblue}{RGB}{50,110,175}
\usepackage[bookmarks=true,colorlinks=true,linkcolor=blue,urlcolor=linkblue,citecolor=blue]{hyperref}
\usepackage{tensor}
\usepackage{graphicx}
\graphicspath{ {./images/} }
\renewcommand\a{\alpha}
\renewcommand\b{\beta}
\renewcommand\d{\delta}
\renewcommand\k{\kappa}
\renewcommand\l{\lambda}
\renewcommand\r{\rho}

\renewcommand\t{\tau}

\newcommand\e{\epsilon}
\newcommand\g{\gamma}
\newcommand\z{\zeta}
\newcommand\m{\mu}
\newcommand\n{\nu}
\newcommand\x{\xi}
\newcommand\p{\pi}
\newcommand\h{\theta}
\newcommand\s{\sigma}
\newcommand\f{\phi}
\newcommand\w{\eta}

\renewcommand\L{\Lambda}

\renewcommand\S{\Sigma}
\renewcommand\O{\Omega}
\renewcommand\H{\Theta}
\newcommand\D{\Delta}

\newcommand{\fig}[1]{Fig.~\ref{#1}}
\newcommand{\eq}[1]{Eq.~(\ref{#1})}

\newcommand{\eqs}[2]{Eqs.~(\ref{#1})-(\ref{#2})}

\newcommand\lb{\left(}
\newcommand\rb{\right)}
\newcommand\ls{\left[}
\newcommand\rs{\right]}
\newcommand\lc{\left\{}
\newcommand\rc{\right\}}
\newcommand{\lan}{\left<}
\newcommand{\ran}{\right>}
\newcommand{\dg}{\dagger}

\newcommand\ra{\rightarrow}

\newcommand{\non}{\nonumber\\}
\newcommand\pt{\partial}
\newcommand{\idp}[2]{\frac{d^{\,#1}{#2}}{(2\p)^#1}}

\newcommand{\eg}{{e.g.}}

\newcommand{\ovy}{\overline{y}}
\newcommand{\ovby}{\overline{\by}}
\newcommand{\cL}{{\cal L}}

\newcommand{\fT}{{\mathfrak{T}}}
\newcommand{\fS}{{\mathfrak{S}}}
\newcommand{\mS}{{\mathcal{S}}}
\newcommand{\mT}{{\mathcal{T}}}
\newcommand{\diag}{{\rm{diag}}}

\newcommand{\LE}{{\rm LE}}
\newcommand{\FO}{{\rm FO}}

\newcommand{\opt}[1]{\widehat{#1}}
\newcommand{\bx}{{\vec x}}

\newcommand{\bp}{{\vec p}}
\newcommand{\bk}{{\vec k}}

\newcommand{\bq}{{\vec q}}

\newcommand{\by}{{\vec y}}

\newcommand{\zero}{{(0)}}
\newcommand{\one}{{(1)}}
\newcommand{\bkn}{{(n)}}
\newcommand{\bkk}{{(k)}}
\newcommand\mc{\mathcal}

\renewcommand{\vec}{\boldsymbol}

\begin{document}
\title{Tensor spin polarization induced by curved freeze-out hypersurface}

\author{Zhong-Hua Zhang}
\affiliation{Physics Department and Center for Field Theory and Particle Physics, Fudan University, Shanghai 200438, China}

\author{Xu-Guang Huang}
\email{huangxuguang@fudan.edu.cn}
\affiliation{Physics Department and Center for Field Theory and Particle Physics, Fudan University, Shanghai 200438, China}
\affiliation{Key Laboratory of Nuclear Physics and Ion-beam Application (MOE), Fudan University, Shanghai 200433, China}
\affiliation{Shanghai Research Center for Theoretical Nuclear Physics, National Natural Science Foundation of China and Fudan University, Shanghai 200438, China}
\begin{abstract}
We investigate how the curvature of the freeze-out hypersurface polarizes massive vector bosons in relativistic heavy-ion collisions. Starting from the Proca Lagrangian and using the Wigner function formalism, we perform a systematic gradient expansion to obtain a covariant spin-polarization tensor expressed in terms of hydrodynamic fields and the curvature tensor of the freeze-out hypersurface. Analytic results for Bjorken and Gubser flows show that curvature anisotropy generates a nonzero tensor polarization. For $\f$ meson, we estimate the curvature contribution to its spin alignment as $ \d\H_{yy} \sim -10^{-4} $ to $ -10^{-3} $. We also find that the curvature contribution grows as the system size decreases. A rough estimate for central O-O collisions gives a spin alignment of order $-10^{-2}$, suggesting that spin-alignment measurements in such small systems may provide a clean probe of this geometric effect.
\end{abstract}
\maketitle

\section{Introduction}
\label{sec:intro}
The discovery of $\L$ hyperons' spin polarization has made spin a new tool to probe the properties of quark-gluon plasma (QGP) in relativistic heavy-ion collisions.
(See Refs.~\cite{Huang:2020xyr,Liu:2020ymh,Huang:2020dtn,Becattini:2020ngo,Becattini:2022zvf,Becattini:2024uha} for reviews.)
The global spin polarization of various hyperons, measured by the STAR Collaboration~\cite{STAR:2017ckg,STAR:2018gyt,STAR:2021beb,STAR:2020xbm}, matches predictions from the spin-thermal vorticity coupling in statistical local equilibrium~\cite{Becattini:2013fla,Fang:2016vpj,Liu:2020flb}.
However, this coupling fails to explain the local spin polarization (azimuthal angle dependence) of $ \L $ as reported by STAR and ALICE Collaborations~\cite{STAR:2019erd,ALICE:2021pzu}.
To reproduce the experimental data, one needs to include other effects such as the spin-thermal shear coupling~\cite{Becattini:2021suc,Becattini:2021iol,Liu:2021uhn,Fu:2021pok,Yi:2021ryh} and other hydrodynamic contributions at leading gradient~\cite{Wu:2019eyi,Liu:2021nyg,Buzzegoli:2021wlg}.
Therefore, investigating spin polarization in QGP produced by heavy-ion collisions requires a closer look at the local equilibrium.

When describing spin polarization in relativistic fluids, researchers often use the local equilibrium density operator~\cite{Zubarev:1979,Weert:1982,Becattini:2019dxo}.
To determine this extended grand canonical ensemble, we need the values of the conserved charges (\eg, energy-momentum density) on a 3D hypersurface.
Most past theoretical works assume the major part of the hypersurface is isochronic~\cite{Becattini:2021suc,Liu:2021nyg,Zhang:2024mhs,Yang:2024fkn} (or nearly isochronic~\cite{Sheng:2024pbw}).
But in numerical simulations, the hypersurface is chosen as the freeze-out hypersurface in order to compare with the experimental data.
Taking the Bjorken expansion as an example~\cite{Bjorken:1982qr}, the freeze-out hypersurface resembles an iso-proper-time surface, not an isochronic one.
Since the thermal current field $ \b_\m(x) $ generally violates the Killing equation $ \pt_{(\m}\b_{\n)} = 0 $ , the spin-thermal shear coupling should depend on the hypersurface. We will further explain this in the following.

Thus, there may appear a novel contribution to the spin polarization due to the geometrical shape of the freeze-out hypersurface. By analyzing the parity and time-reversal properties of the gradient expansion~\cite{Zhang:2024mhs,Sheng:2024pbw}, one can verify that the vector spin polarization at leading order (the terms proportional to thermal vorticity, thermal shear, etc.) would not be affected by the curvature of the hypersurface.
The effect of the curvature would appear at the next-to-leading order for vector polarization. (See a recent study of this issue for fermions in Ref.~\cite{Sheng:2025cjk}.) To highlight how the shape of hypersurfaces matters, we study the tensor spin polarization of massive vector bosons, which, as we will show, would be affected by the curvature at the leading order.

For the spin polarization of vector mesons, the STAR Collaboration measured the global spin alignment of $ \f $~\cite{STAR:2022fan}, and ALICE observed the $ p_T $ dependent spin alignment of $ J/\psi $ and $ D^{*+} $~\cite{ALICE:2020iev,ALICE:2022dyy,ALICE:2025cdf}.
These are all tensor polarizations. Understanding these experimental results might require introduction of some novel mechanisms, such as strong $\phi$ mean field~\cite{Sheng:2019kmk,Sheng:2022wsy,Sheng:2023urn}, strong gluonic fields~\cite{Kumar:2023ghs,Yang:2024qpy}, critical fluctuations~\cite{Chen:2024hki}, and others~\cite{Sheng:2024kgg,Xu:2024kdh,Ahmed:2025bwi,Liang:2025hxw,Zhu:2025rdj,Sahoo:2025bkx,Yan:2025tlx}. However, as a benchmark of the spin polarization, it is important to first clarify the contributions of hydrodynamic fields ~\cite{Zhang:2024mhs,Yang:2024fkn}. Previous studies adopted the ansatz that the freeze-out hypersurface is mostly a hyperplane except for the timelike edge~\cite{Zhang:2024mhs,Yang:2024fkn}, thus missing the possible geometric effects due to the nontrivial shape of the freeze-out hypersurface. In this paper, we demonstrate that the tensor spin polarization induced by the curvature of the hypersurface is at the first order in gradient in hydrodynamic fields, which are one order lower than the ones from purely hydrodynamic fields~\cite{Zhang:2024mhs,Yang:2024fkn}.

The paper is organized as follows.
In Sec.~\ref{sec:LEstate}, we discuss the description of the spin state of massive bosons using covariant vector and tensor.
In Sec.~\ref {sec:density-operator}, we introduce the local equilibrium density operator for spinful fluids and perform the cumulant expansion.
In Sec.~\ref{sec:wigner}, we evaluate the Wigner function under the density operator with Feynman rules.
In Secs.~\ref{sec:Bjorken} and~\ref{sec:Gubser}, we simplify the polarization tensor in Bjorken and Gubser flows to estimate the magnitude.
We summarize our main result and discuss it briefly in Sec.~\ref{sec:conclusion}.

Within this paper, we use the Minkowski metric $ \w_{\m\n} = \diag(1, -1, -1, -1) $ and adopt natural units $ k_B = c = \hbar =1 $.
In Minkowski spacetime, the Levi-Civita tensor $ \e^{\m\n\r\s} $ is normalized as $ \e^{0123} = 1 $;
in three spacelike dimensions, $ \e^{ijk} $ is normalized as $ \e^{123} = 1 $.
We define the shorthand notations $ A^{(\m\n)} = (A^{\m\n} +A^{\n\m}) /2 $ and $ A^{[\m\n]} = (A^{\m\n} -A^{\n\m}) /2 $ for an arbitrary rank-2 tensor $ A^{\m\n} $ to represent its symmetric and antisymmetric parts.

\section{Spin State of a Massive Vector Boson}
\label{sec:LEstate}
Let us briefly discuss how to describe the spin state of a massive vector boson; more information can be found in Ref.~\cite{Zhang:2024mhs}. For neutral massive vector bosons,\footnote{We focus on neutral vector bosons in the paper, but all the discussions can be straightforwardly extended to charged vector bosons.} such as $ \f $ or $ J/\psi$ mesons, their dynamics is governed by the Proca Lagrangian 
\begin{equation}\label{def-Lagrangian}
    \cL = -\frac{1}{4} F_{\m\n}F^{\m\n} +\frac{1}{2} m^2 A_\m A^\m
\end{equation}
with $ F_{\m\n} \equiv \pt_\m A_\n - \pt_\n A_\m $ the field tensor, $ m $ the mass of the free vector bosons. The corresponding equations of motion are the on-shell condition, $ (\pt^2+m^2)A_\m = 0 $, and the Lorentz condition, $ \pt_\m A^\m = 0 $.
After quantization, the vector field operator can be decomposed as
\begin{equation}\label{def-field_decomp}
    \opt{A}^\m (x) = \sum_{s=1}^3\int\idp{3}{\bk} \frac{1}{2E_\bk} \ls \opt{a}_\bk^s \e^\m_s(k)e^{-ik\cdot x} + \opt{a}_\bk^{\dg s} \e^{\m*}_{s}(k) e^{ik\cdot x}\rs
\end{equation}
with $ k^0 = E_\bk = \sqrt{\bk^2+m^2} $ following the on-shell condition,
$ \e_s(k) $ the spin polarization vectors,
and $ \opt{a}_\bk^s $ and $ \opt{a}_\bk^{\dg s} $ the annihilators and creators for a vector boson with momentum $ \bk $ and polarization index $ s $.
The annihilators and creators follow the commutation relation
\begin{equation}\label{commutator}
    [\opt{a}_\bk^s,\opt{a}_{\bk'}^{\dg s'}] = (2\p)^3 (2E_\bk) \d^{(3)}(\bk-\bk') \d^{ss'}\,.
\end{equation}
The polarization vector $ \e_s(k) $ is defined by $ \e^\m_s(k) = [k]\indices{^\m_\n} e^\n_s $, where $ e^\n_s = \w\indices{^\n_s} $ and $[k]$ is a standard Lorentz transformation from the rest frame of the particle to the frame in which the particle's four-momentum is $k$. For simplicity, we choose the standard Lorentz transformation $ [k] $ as a pure Lorentz boost $ L(k) $
defined by
\[
    L\indices{^0_0}(k) = E_\bk/m\,, \qquad
    L\indices{^0_i}(k) = L^i_{~0}(k) = k^i/m\,, \qquad
    L\indices{^i_j}(k) = \d^{i}_j - k^i k_j/[m(E_\bk+m)]\,.
\]
Therefore, we can express the spin polarization vector in an explicit way
\begin{equation*}
    \e^\m_s(k) = e^\m_{s} -\frac{k\cdot e_{s}}{m} \frac{k^\m + m~e_0^\m}{e_0\cdot k+m},\quad \text{with } s = 1,2,3 \,,
\end{equation*}
where $ e_0^\m = \w^{\m}_{~~0} $ denotes the time direction.
Based on the above definition, we have the following completion and orthonormal relations for the polarization vectors
\begin{align}
\label{compeltion-1}
    \sum_{s=1}^{3} \e_s^\m(k) \e_s^\n(k) =&~ - \D_\bkk^{\m\n} \equiv -\w^{\m\n} +\frac{k^\m k^\n}{m^2},\\
\label{compeltion-2}
    \e_s^\m(k) \e_r^\n(k) \w_{\m\n} =&~ -\d_{rs}\,,
\end{align}
which show that three spacelike polarization vectors form a complete orthonormal basis of the subspace orthogonal to $ k $.

We describe the spin polarization state by the so-called spin density matrix.
In a relativistic quantum field theory, this matrix can be defined as \cite{Becattini:2020sww}
\begin{equation} \label{spin-density-matrix}
    \H_{rs}(k) = \frac{\Tr \lb \opt{\r}\,\opt{a}^{s\dg}_{\bk} \opt{a}^{r}_{\bk}\rb}
    {\sum_t \Tr \lb \opt{\r}\, \opt{a}^{t\dg}_{\bk} \opt{a}^{t}_{\bk} \rb}\,,
\end{equation}
where $k$ is the four-momentum of the particle in the laboratory frame. 
We choose $ r $ and $ s $ to be $ 1,\,2,\,3 $,
corresponding to $ x,\,y,\,z $, the indices of adjoint representation.
Spin density matrix $ \H(k) $ is a $ 3 \times 3 $ Hermitian matrix with unit trace, which can be decomposed as
\begin{equation}\label{density-matrix-decomp}
    \H(k) = \frac{1}{3} I + \frac{1}{2} \sum_i \fS^i(k) s^i + \sum_{ij} \fT^{ij}(k) \S^{ij}\,,
\end{equation}
where $ (s^i)_{jk} \equiv -i \e^{ijk} $, the adjoint representation of $ SO(3) $ fulfilling the commutation relation $ [s^i, s^j] = i \sum_k \e^{ijk} s^k $,
and $ \S^{ij} \equiv s^{(i} s^{j)} - \vec{s}^2 \d^{ij}/3 $.
We can contract the polarization vector and tensor from the spin density matrix
\begin{align}
\label{vector-polar}
    \fS^i(k) =&~ \tr \lc s^i \H(k) \rc = i \sum_{jk} \e^{ijk} \H_{jk}(k) \,, \\
\label{tensor-polar}
    \fT^{ij}(k) =&~ \tr \lc \S^{ij} \H(k) \rc = -\H_{(ij)}(k) +\frac{1}{3}\d^{ij} \,.
\end{align}

We should point out that the exact forms of $ \H_{rs}(k) $ and $ \e_s(k) $ depend on the choice of the standard Lorentz transformation~\cite{Zhang:2024mhs}.
To avoid this, we define the Lorentz covariant spin polarization vector and polarization tensor
\begin{align}
\label{covariant-polar-vector}
    \mS^\m(k) \equiv&~ \sum_i \fS^i (k) \e^\m_i(k)\,,\\
\label{covariant-polar-tensor}
    \mT^{\m\n}(k) \equiv&~ \sum_{ij} \fT^{ij} (k) \e^\m_i(k)\e^\n_j(k)\,.  
\end{align}
Considering the transformation of $ \H_{rs}(k) $ and $ \e_s(k) $ under the little group preserving momentum $ k $, it is straightforward to prove that the covariant spin polarization vector and tensor are independent of the choice of the standard Lorentz transform. (See Ref.~\cite{Zhang:2024mhs}.)
It has been proved in Ref.~\cite{Becattini:2020sww} (for the single-particle case) that
$ \mS^\m (k) $ in \eq{covariant-polar-vector} is the expectation value of the operator defined by the Pauli-Lubanski vector $ \opt{W}^\m $
\begin{equation}\label{def-spin_vector_operator}
    \opt{\mS}^\m \equiv -\frac{1}{m} \opt{W}^\m = -\frac{1}{2m} \e^{\m\n\r\s} \opt{J}_{\n\r} \opt{P}_\s\,,
\end{equation}
where $ \opt{J}_{\n\r} $ represents the angular momentum operator and $ \opt{P}_\s $ the four-momentum operator.
Similarly, we demonstrate in Appendix \ref{appx:covariant-tensor} that $ \mT^{\m\n}(k) $ defined by \eq{covariant-polar-tensor} is the expectation value of the tensor operator 
\begin{equation}\label{def-spin_tensor_operator}
    \opt{\mT}^{\m\n} \equiv \opt{\mS}^{(\m} \opt{\mS}^{\n)} +\frac{2}{3}\lb \w^{\m\n} -\frac{1}{m^2} \opt{P}^\m \opt{P}^\n\rb \,,
\end{equation}
which is given by Ref.~\cite{Leader:2011vwq}.
Equaiton \eqref{tensor-polar} shows that the spin alignment along a chosen quantization axis can be directly obtained from the tensor polarization. In particular, for a spatial axis $a=x,y,z$, we express the spin alignment by
\begin{equation}
    \d\H_{aa}(k) \equiv \H_{aa}(k)-\frac{1}{3} = -\,\fT^{aa}(k) \,.
\end{equation}
In particular, the spin alignment measured in experiments~\cite{STAR:2022fan,ALICE:2020iev,ALICE:2022dyy,ALICE:2025cdf} is $ \d\H_{yy}(k) $. Using the covariant decomposition in \eq{covariant-polar-tensor}, this can be rewritten as
\begin{equation}\label{def-spin_align}
\delta\Theta_{aa}(k)
= -\,\epsilon_a^\mu(k)\epsilon_a^\nu(k)\,\mT_{\mu\nu}(k) \,.
\end{equation}

To calculate particle polarization in a relativistic fluid, it is helpful to introduce the Wigner operator
\begin{equation}
\label{def-Wigner_funciton}
    \opt{W}^{\m\n}(x,k) = \int d^4 y\,e^{ik\cdot y} \opt{A}^{\n}\left( x-\frac{y}{2} \right) \opt{A}^{\m} \left(x+\frac{y}{2} \right) \h(k^2)\h(k^0) \,,
\end{equation}
where the Heaviside step function $ \h $ appears as we take the particlelike branch of the Wigner operator with positive energy.
By plugging in the decomposition of the field operator shown in \eq{def-field_decomp}, we obtain
\begin{equation}\label{Wigner-decomposition}
\begin{split}
    \opt{W}^{\m\n}(x,k) =
    &~ \int\idp{3}{\bp_1} \idp{3}{\bp_2}\frac{(2\p)^4}{(2E_{\bp_1})(2E_{\bp_2})} \sum_{r_1 \, r_2} e^{-i(p_1-p_2)\cdot x} \\
    &~ \times \d^{4}\lb k-\frac{p_1}{2}-\frac{p_2}{2}\rb \opt{a}^{r_2\dg}_{\bp_2}\opt{a}^{r_1}_{\bp_1} \e^{\m}_{r_1}(p_1)\e^{\n*}_{r_2}(p_2) \,.
\end{split}
\end{equation}
From the above expression, the Wigner operator fulfills the Boltzmann equation~\cite{Becattini:2020sww}
\begin{equation}
\label{eq-Boltzmann}
    k^\a \frac{\pt}{\pt x^\a} \opt{W}^{\m\n}(x,k) = 0 \,
\end{equation}
as $ p_1 $ and $ p_2 $ lie on the mass shell during the integration and therefore $ (p_1 + p_2) \cdot (p_1 - p_2) = 0 $.
Using \eq{eq-Boltzmann}, we can extract a pair of creator and annihilator for the Wigner operator by integration over a 3D hypersurface with proper boundary conditions~\cite{Becattini:2020sww}
\[
    \d(k^2-m^2) \h(k^0) a_\bk^{s\dagger} a_\bk^{r} = \frac{1}{\p} \e^\m_r(k) \, \e^\n_s(k) \int d\S \cdot k \, \opt{W}_{\m\n} (x,k) \,.
\]
Therefore, the density matrix defined in \eq{spin-density-matrix} can be expressed by a Cooper-Frye type formula~\cite{Becattini:2020sww,Zhang:2024mhs}
\begin{equation}\label{cooper-frye_spin}
    \H_{rs} (k) = \frac{ \e^\m_r(k) \, \e^\n_s(k) \int_{\S} d\S\cdot k \, W_{\m\n}(x,k)}{ \sum_{t} \e^\m_t(k) \, \e^\n_t(k) \int_{\S} d\S\cdot k \, W_{\m\n}(x,k)}
\end{equation}
with the help of the Wigner function, which is defined by the expectation value of the Wigner operator
\begin{equation}
    W^{\m\n}(x,k) \equiv \Tr{\opt{\r} \, \opt{W}^{\m\n}(x,k)} \,.
\end{equation}
In general, the argument $ k $ in the Wigner function is not necessarily on the mass shell,
however, in \eq{cooper-frye_spin}, the integration over $ \S $ limits it to be on shell.

In the following calculation, we focus on the projected Wigner function, which can be decomposed into scalar, vector, and tensor components
\begin{equation}
    W_\perp^{\m\n} (x,k) \equiv \D_\bkk^{\m\a} \D_\bkk^{\n\b} W_{\a\b}(x,k)
    = (2\p) f(x,k) \lc -\frac{1}{3} \D^{\m\n}_\bkk + \frac{i}{2m} \e^{\m\n\r\s} k_\r \mathbb{S}_\s(x,k) - \mathbb{T}^{\m\n} (x,k) \rc \,.
\end{equation}
One may extract them from the Wigner function by
\begin{align}
\label{extract-scalar}
    f(x,k) =&~ -\frac{1}{2\p} \D_\bkk^{\m\n} W_{\m\n}(x,k) \,,\\
\label{extract-vector}
    \mathbb{S}^{\m}(x,k) =&~ -\frac{i}{(2\p) f(x,k) m} \e^{\m\n\a\b}k_\n W_{\a\b}(x,k) \,,\\
\label{extract-tensor}
    \mathbb{T}^{\m\n}(x,k) =&~ -\frac{1}{(2\p) f(x,k)} W^{\lan\m\n\ran}_{\perp}(x,k) \,,
\end{align}
where $ W_\perp^{\lan\m\n\ran}(x,k) $ denotes the symmetric and traceless part of the projected Wigner function 
$ W^{\lan\m\n\ran}_{\perp} = \D_\bkk^{\m\a} \D_\bkk^{\n\b} W_{(\a\b)}(x,k) - \D_\bkk^{\m\n} \D_\bkk^{\a\b} W_{\a\b}(x,k) / 3 $. 
Plugging \eq{cooper-frye_spin} into \eq{vector-polar} and using \eq{covariant-polar-vector}, we obtain
\begin{equation}\label{cooper-frye-vector}
    \mS^\m(k) = \frac{\int d\S\cdot k \, f(x,k) \, \mathbb{S}^{\m}(x,k)}{\int d\S\cdot k \, f(x,k)} \,,
\end{equation}
where we have used the expression $ \sum_{ijk} \e^{ijk} \e^{\m}_i \e^{\r}_j \e^{\s}_k = \e^{\n\m\r\s} k_\n /m $.
Similarly, we can use the tensor component of the Wigner function to express the spin polarization tensor in a Cooper-Frye type formula
\begin{equation}\label{cooper-frye-tensor}
    \mT^{\m\n}(k) = \frac{\int d\S\cdot k \, f(x,k) \, \mathbb{T}^{\m\n}(x,k)}{\int d\S\cdot k \, f(x,k)} \,,
\end{equation}
to derive which we substitute \eq{cooper-frye_spin} to \eq{tensor-polar} and use \eq{covariant-polar-tensor}.
Inspired by \eq{def-spin_align}, we introduce the phase-space spin alignment
\begin{equation}
    \d\H_{aa}(x,k) \equiv -\,\e_a^\m(k)\e_a^\n(k)\,\mathbb{T}_{\m\n}(x,k) \,.
\end{equation}
While the observable $ \d\H_{aa}(k) $ is obtained only after integrating over the freeze-out hypersurface
\begin{equation}
    \d\H_{aa}(k) = \frac{\int d\S\cdot k \, f(x,k) \, \d\H_{aa}(x,k)}{\int d\S\cdot k \, f(x,k)} \,.
\end{equation}
Therefore, once $\mathbb{T}_{\mu\nu}(x,k)$ is known, one can directly infer the corresponding phase-space spin alignment and then compare with the experimental observable after the freeze-out average.

\section{Local Equilibrium Density Operator and its Expansion}
\label{sec:density-operator}
For a relativistic fluid that has reached local thermal equilibrium, we express its density operator by~\cite{Zubarev:1979,Weert:1982,Becattini:2019dxo}
\begin{equation}\label{LEDO}
    \opt{\r}_\LE = \frac{1}{Z_\LE}\exp\bigg\{-\int_{\S} d\S_\m(y) \ls \opt{T}^{\m\n}(y)\b_\n(y) -\frac{1}{2}\opt{S}^{\m\r\s}(y)\O_{\r\s}(y) \rs \bigg\} \,,
\end{equation}
where $ Z_\LE $ is the partition function playing the role of normalization factor ensuring $ \Tr \opt{\r}_\LE = 1 $,
$ d\S_\m $ denotes $ d\S ~n_\m $ with $ d\S $ the measure of the hypersurface $ \S$ and $ n_\m $ the normal vector,
$ \b_\n = u_\n / T $ represents the thermal current,
and $ \O_{\r\s} $ is the spin potential.
In this paper, we treat the spin potential as an independent hydrodynamic field that manipulates the spin of particles in the relativistic fluid described by spin hydrodynamics~\cite{Florkowski:2017ruc,Hattori:2019lfp,Fukushima:2020ucl,Hongo:2021ona,Cao:2022aku,She:2021lhe,Gallegos:2021bzp,Huang:2024ffg}.
Since the density operator at local equilibrium depends on the choice of pseudo-gauge~\cite{Becattini:2018duy,Liu:2021nyg,Buzzegoli:2021wlg},
we choose the energy-momentum tensor and spin tensor in \eq{LEDO} to be the canonical currents defined by
\begin{align*}
    \opt{T}^{\m\n} &= - \opt{F}^{\m\r}\pt^\n \opt{A}_\r-\w^{\m\n}\left(-\frac{1}{4} \opt{F}_{\r\s}\opt{F}^{\r\s}+\frac{1}{2}m^2\opt{A}_\r \opt{A}^\r\right) \,,\\
    \opt{S}^{\m\r\s} &= -\opt{F}^{\m\r}\opt{A}^{\s}+ \opt{F}^{\m\s}\opt{A}^{\r} \,.
\end{align*}

We should emphasize that the density operator defined by \eq{LEDO} depends on the choice of the 3D hypersurface $ \S $.
Typically, it is chosen to be $ \S_{\text{eq}} $, where the fluid truly reaches local thermodynamic equilibrium.
The difference between the density operators at $\S_{\text{eq}} $ and at $\S$ comes from the dissipative effects, which should be considered when dealing with a realistic fluid (\eg~QGP produced by relativistic heavy ion collision).
As for the spin, dissipative effects also contribute to spin polarization~\cite{Li:2022vmb,Buzzegoli:2025zud}.
However, this paper only focuses on the nondissipative contributions to spin polarization of particles at $ \S_\FO $, where spin freezes out.
Therefore, in the following, we choose the $ \S=\S_\FO $ as the hypersurface in \eq{LEDO}.

When evaluating the expectation value of a local operator in the fluid, the gradient expansion has been frequently used when dealing with spin polarization
~\cite{Becattini:2013fla,Becattini:2021suc,Buzzegoli:2022fxu,Liu:2021nyg,Buzzegoli:2021wlg,Sheng:2024pbw,Zhang:2024mhs,Yang:2024fkn}.
Specifically, we evaluate the ensemble average by
\begin{equation}\label{expectation-O}
    O(x) = \Tr{\opt{\r}_\LE \opt{O}(x)} = \frac{1}{Z_\LE} \Tr{e^{\opt{A}+\opt{B}}\opt{O}(x)} \,,
\end{equation}
where we introduce
\begin{align}
\label{def-A}
    \opt{A} &= -\opt{P}^\m\b_\m(x) \,,\\
\label{def-B}
    \opt{B} &= -\int_{\S_{\FO}} d\S_\m(y) \ls \opt{T}^{\m\n}(y)\D\b_\n(y)-\frac{1}{2}\opt{S}^{\m\r\s}(y)\O_{\r\s}(y)\rs
\end{align}
with $\opt{P}^\m=\int_{\S_{\FO}} d\S_\n \opt{T}^{\n\m} $ the momentum tensor and $ \D\b_\m(y)=\b_\m(y)-\b_\m(x) $.
We adopt the power-counting rules by $ \D\b_\m(y) \sim \O_{\r\s}(y) \sim \mc{O}(\pt) $,
which is implied by the magnitude of hyperon polarization in the heavy ion collision experiments~\cite{STAR:2017ckg,STAR:2018gyt,STAR:2021beb,STAR:2020xbm}.
Based on the rules, we expand the exponential expression up to the first order of the gradient
\[
    e^{\opt{A}+\opt{B}} = e^{\opt{A}}\lb 1 + \int_0^1 d\l \, e^{-\l \opt{A}} \opt{B} e^{\l \opt{A}} +\mc{O}(\pt)^2 \rb \,.
\]
Substituting the above expression into \eq{expectation-O}, we obtain
\begin{equation}
    O(x) = O^\zero(x) + O^\one(x) + \mc{O}(\pt)^2 \,,
\end{equation}
where
\begin{align}
\label{exp-O0}
    O^\zero(x) &= \lan \opt{O}(x) \ran_0 \equiv \Tr{ e^{\opt{A}}\opt{O}(x)} \Big/ \Tr e^{\opt{A}} \,, \\
\label{exp-O1}    
    O^\one(x) &= \lan \opt{B}_1 \opt{O}(x) \ran_{0,c}
\end{align}
with
{\small
\[
    \opt{B}_1 = \int_0^1 d\l \int_{\S} d\S_\m(y) \ls -\pt_\a\b_\n(x) (y-x)^\a \opt{T}^{\m\n}(y-i\l\b(x)) + \frac{1}{2}\O_{\r\s}(x)\opt{S}^{\m\r\s}(y-i\l\b(x))\rs
\]
}
and {\footnotesize $ \lan \opt{B}_1 \opt{O}(x) \ran_{0,c} = \lan \opt{B}_1 \opt{O}(x) \ran_{0} - \lan \opt{B}_1 \ran_{0} \lan \opt{O}(x) \ran_{0} $} the connected part of the correlator.
These expressions imply that the gradient expansion is valid, namely {\small $ \abs{O^\one(x)} \ll \abs{O^\zero(x)} $},
only if the coefficients in \eq{def-B} are small enough within the correlated region of the operators.
Let us take the thermal current as an example. The Taylor expansion of $ \b_\n(y) $ around $ x $ works only if $ \abs{\b}/\abs{\pt\b} \gg l $,
where $ \abs{\b}/\abs{\pt\b} $ stands for the typical length of the variation of the thermal current and $ l $ stands for the correlation length of $ \opt{T}^{\m\n}(y-i\l\b(x)) $ and $ \opt{O}(x) $.

When evaluating $ O^\one(x) $, we need to calculate an integral over the hypersurface $ \S $.
However, we only need to consider a small region around $ x $ during the integration, since \eq{exp-O1} contains only the connected correlation,
whose contribution can be ignored in the region far beyond the correlation length.
Such a claim allows us to perform the integral over an imaginary hypersurface whose point is close to the real hypersurface within the neighborhood of $ x $.

We assume the 3D hypersurface $ \S $ can be described as the zero set of an analytic function $ F(x) $
\begin{equation}\label{def-F}
    F(x) = 0 \,, \quad \forall x \in \S \,. 
\end{equation}
In the following, we consider point $ x $ belonging to the spacelike part of the surface.
By the definition of ``spacelike,'' we have
\[
    \frac{\pt F(x)}{\pt x^\m}\frac{\pt F(x)}{\pt x_\m} > 0 \,.
\]
Therefore, we can define the timelike normal vector by $n^\mu=v^\mu/v$, where $ v_\m(x) = \pt F(x)/\pt x^\m $
and $ v = \sqrt{v\cdot v} $. We also choose the orientation such that $n^0>0$; if this condition is not met,
one may simply replace $F(x)$ by $-F(x)$.
When evaluating $ O^\one(x) $, the integration over $\S$ receives relevant contributions only from a small neighborhood
of $x$ within the correlation length $ l $. Therefore, it is sufficient to approximate the hypersurface locally around $x$.
For this purpose, we choose a local Lorentz frame at $x$ such that $ n^\mu(x)=(1,\vec{0}) $.
In this frame, since $\Sigma$ is spacelike at $x$, one has $ \pt_0 F(x) \neq 0 $,
and thus by the implicit function theorem the hypersurface can be solved locally as
\[
    y^0 = \phi(\vec{y}) \,, \quad
    F(\phi(\vec{y}),\vec{y}) = 0
\]
in a neighborhood of $x$. Differentiating $ F(\f(\by),\by)=0 $ with respect to $ y^i $, we obtain
\begin{equation}\label{eq:diff-F}
    0 = \left. \frac{\pt F(y)}{\pt y^0} \, \frac{\pt\f(\by)}{\pt y^i} + \frac{\pt F(y)}{\pt y^i} \right|_{y^0 = \f(\by)} \,.
\end{equation}
In this frame, since the tangent plane of $\Sigma$ at $x$ is orthogonal to $(1,\vec{0})$,
the first spatial derivatives of $\phi$ vanish at $x$, namely, $ 0 =\left.\pt\f/\pt y^i\right|_{\by=\bx} $.
Based on \eq{eq:diff-F}, this expression implies $ \partial_iF(x)=0 $ in the rest frame of $ n^\mu(x) $.
Differentiating \eq{eq:diff-F} once more with respect to $y^j$, we obtain
\[
    0 = \left. \frac{\pt F}{\pt y^0}\,\frac{\partial^2\phi}{\partial y^i\partial y^j}
    +\frac{\pt^2 F}{\pt y^0\pt y^0}\,\frac{\partial\phi}{\partial y^i}\frac{\partial\phi}{\partial y^j}
    +\frac{\pt^2 F}{\pt y^0\pt y^i}\,\frac{\partial\phi}{\partial y^j}
    +\frac{\pt^2 F}{\pt y^0\pt y^j}\,\frac{\partial\phi}{\partial y^i}
    +\frac{\pt^2 F}{\pt y^i\pt y^j} \right|_{y^0 = \f(\by)} \,.
\]
Evaluating this equation at $y=x$, all terms containing first derivatives of $\phi$ vanish, and thus
\[
    \pt_i\pt_j \f(\bx) = - \frac{\pt_i\pt_j F(x)}{\pt_0 F(x)} \,.
\]
Since in the chosen frame $ n \cdot (y-x) = y^0-x^0 $, the Taylor expansion yields
\begin{equation}\label{taylor:surface-local}
    n \cdot (y-x) = \phi(\mathbf y)-\phi(\mathbf x) 
    = - \frac{1}{2} \, \frac{\pt_i\pt_j F(x)}{\pt_0 F(x)} (y-x)^i (y-x)^j + \mc{O}\big( |\by-\bx|^3 \big) \,.
\end{equation}
Expressing the above result in a covariant form, we obtain a relation that holds in any reference frame
\begin{equation}\label{taylor:surface}
    n \cdot (y-x) = \frac{1}{2} B_{\m\n} (y-x)^\m (y-x)^\n + \mc{O}\big( |y-x|_\perp^3 \big) \,,
\end{equation}
where $|y-x|_\perp$ denotes the spatial distance between $x$ and $y$ measured within the tangent hyperplane, namely, {\footnotesize $ \lb-\D_\bkn^{\m\n}(y-x)_\m(y-x)_\n\rb^{1/2} $ }.
\footnote{$\D^\bkn$ is the projection tensor defined by $ \D^\bkn_{\m\n} = (\w_{\m\n}-n_\m n_\n) $}
Here $ B^{\m\n}(x) $ is the curvature tensor \footnote{Curvature tensor $ B $ depends on how the 3D hypersurface is embedded in the higher-dimensional space, in this paper, the Minkowski space-time.
It differs from an intrinsic curvature tensor like the Riemann curvature tensor.} of $ \S $ defined by
\begin{equation}\label{def-curv}
    B^{\m\n}(x) \equiv -\D_\bkn^{\m\a}\frac{\pt n^\n(x)}{\pt x^\a} = -\frac{1}{v} \D_\bkn^{\m\r}\D_\bkn^{\n\s} \frac{\pt^2 F(x)}{\pt x^\r \pt x^\s} \,.
\end{equation}
To derive the second equality, we use
\[
    \pt_\a n^\n(x) = \frac{1}{v} \ls \pt_\a \pt^\n F(x) - n^\n \, \pt_\a v(x) \rs \,.
\]
Moreover, from the normalization condition $n_\nu n^\nu=1$, one finds $ n_\n \pt_\a n^\n(x) = 0 $, which implies that $\partial_\alpha n^\nu(x)$ is tangent to $\Sigma$ with respect to the index $\nu$.
Therefore,
\[
    \pt_\a n^\n(x) = \D_{(n)}^{\n\s}\pt_\a n_\s(x) \,.
\]
Using this relation, the second equality in \eqref{def-curv} follows immediately.
It is straightforward to verify that \eq{taylor:surface} reduces to \eq{taylor:surface-local}
in the frame of $ n^\mu(x)=(1,\vec{0}) $, by noticing that $ \D_{\m\r}^\bkn(x) \pt^\r \rightarrow (0,\vec{\pt}) $ and $ v \rightarrow \pt_0 F(x) $.
Equation~\eqref{taylor:surface} shows that, up to quadratic order, the normal displacement of the hypersurface
away from its tangent plane is fully characterized by $B_{\mu\nu}(x)$.
We can consider a virtual hypersurface $ \S^\prime $ determined by
\[
    n \cdot (y'-x) = \frac{1}{2} B_{\m\n}(x) (y'-x)^\m (y'-x)^\n \,, \quad \forall y' \in \S^\prime \,,
\]
to replace the true hypersurface $\S$ as long as they are close enough within the correlated region. 
See \fig{fig:variable_substitution} for a sketch.
To be precise, we are assuming
\[
    |n \cdot (y' -y)| \ll l \,, \quad \text{for } \overline{y}'_\n = \overline{y}_\n \,,
\]
within the neighborhood of $ x $, $ |y-x|_\perp \sim l $.
Here, $ y' \in \S' $, $ y \in \S $. We define $ \overline{y} $ as the projection of $ y $ onto the tangent plane $ P(x) $, given by
\[
    \overline{y}^\m = x^\m + \D^{\m\n}_{(n)} (y-x)_\n \,,
\]
while $ \overline{y}' $ denotes the projection of $ y' $ onto $ P(x) $.
For an analytic hypersurface, this should be true as long as $ l $ is small enough.
In the following, we take this assumption and do not distinguish between $ \S $ and $ \S' $.
\section{Wigner Function and Spin Polarization at local equilibrium}
\label{sec:wigner}
To evaluate the Wigner operator, we use the expansion discussed in the previous section.
Using the commutation relation \eq{commutator}, it is straightforward to obtain
\[
    \lan \opt{a}^{r\dg}_\bk\opt{a}^{s}_\bq\ran_0 = (2E_\bk)(2\p)^3\d^{(3)}(\bk-\bq) \d^{rs} n_B(\b(x)\cdot k) \,.
\]
Decomposing the Wigner operator into creator and annihilator [see \eq{Wigner-decomposition}], we derive
\begin{equation}\label{result-Wigner-0}
    W^{\m\n}_\zero(x,k) = -(2\p)\h(k^0) \d(k^2-m^2) \D^{\m\n}_\bkk n_B(\b(x)\cdot k) \,,
\end{equation}
where $ n_B(\b(x)\cdot k) $ denotes the Bose-Einstein distribution in the rest frame of $ \b_\m(x) $, namely, $ 1/(e^{\b(x)\cdot k}-1) $.
Therefore, the scalar distribution reads
\begin{equation}
    f(x,k) =3 \d(k^2-m^2) \h(k^0)  n_B(\b(x)\cdot k) + \mc{O}(\pt)
\end{equation}
at the leading order.

In the following, we demonstrate the calculation under the scenario that $ n_\m(x) = (1,\vec{0}) $.
Making such an assumption will simplify our demonstration while keeping its generality.
This is because both the Wigner operator and the element in the density operator are Lorentz covariant.
Therefore, if $ n_\m(x) $ does not meet the criteria, one may calculate the Wigner function in a frame where $ n_\m(x) = (1,\vec{0}) $,
then perform a Lorentz transformation to obtain the final result in the original global frame.

\begin{figure}
    \centering
    \includegraphics[width=0.65\linewidth]{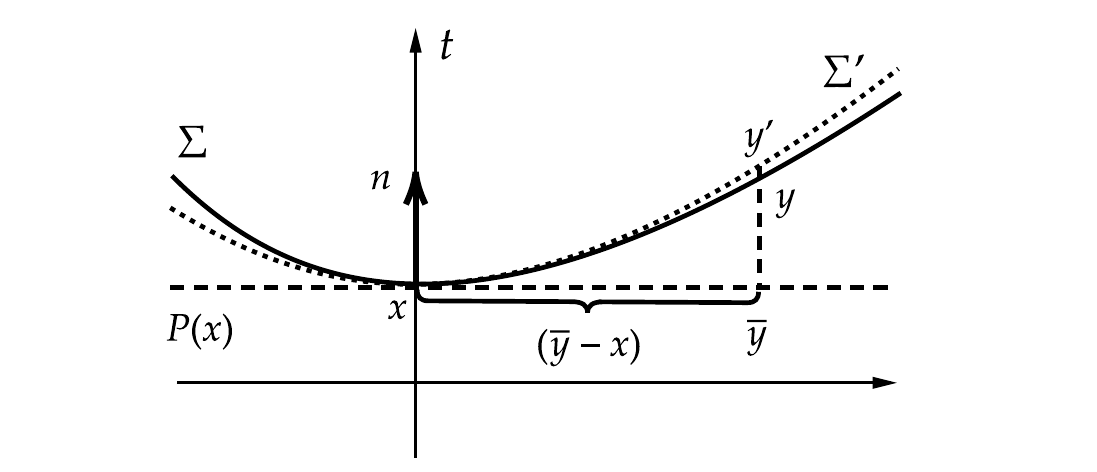}
    \caption{Sketch of $ \S $ and $ \S'$ in the local rest frame of $ n_\m(x) $.
    $ P(x) $ stands for the tangent plane of the hypersurface at point $ x $.
    $ y $ is a point on $ \S $; $ y' $ is a point  on $ \S' $; $ \ovy $ is a point on the tangent plane.
    Since $ \S $ and $ \S' $ are close enough within the correlated region, we do not distinguish between $ \S $ and $ \S' $ in the following.
    During the variable substitution, $ \ovy(y) $ is chosen to be a function of $ y $ so that $ (y-\ovy)_\m \propto n_\m $.}
    \label{fig:variable_substitution}
\end{figure}

Now, let us consider the first-order result.
To avoid repetition, we only exhibit the calculation of the first part of \eq{exp-O1}, the part induced by the energy-momentum tensor, that is
\begin{equation}
    W^{\m\n}_\one\big|_T (x,k) = -\pt_\a\b_\s(x) \int_0^1 d\l \int_{\S} d\S_\r(y)\, (y-x)^\a \lan \opt{T}^{\r\s}(y-i\l\b(x)) \opt{W}^{\m\n}(x,k)\ran_{0,c} \,.
\end{equation}
To evaluate the integration over $ \S $, we perform the variable substitution $ y \to \overline{y} $ and rewrite the integral as an integration over the tangent plane $ P(x) $, as illustrated in \fig{fig:variable_substitution}.
Accordingly, we make the replacements
\begin{align*}
    \int_\S d\S_\m(y) &\ra \int_{P(x)} d^3 \ovby \ls n_\m(x)-B_{\m\n}(x)(\ovy-x)^\n\rs \,,\\
    (y-x)^\a &\ra (\ovy-x)^\a + n\cdot (y-\ovy) n^\a \,,\\
    \opt{T}^{\m\n}(y) &\ra \opt{T}^{\m\n}(\ovy) + n\cdot (y-\ovy) (n\cdot \pt)\opt{T}^{\m\n}(\ovy) + \cdots
\end{align*}
with
\[
    n\cdot(y-\ovy) = n\cdot(y-x) = \frac{1}{2} B_{\m\n} (\ovy-x)^\m(\ovy-x)^\n \,.
\]
During the calculation, we disregard the items represented by the ellipsis, since we assume that $ \abs{n\cdot (y-x)} $ is small within the correlated region of operators.
Expressly, we assume the curvature of the hypersurface is slight enough to satisfy $ \abs{\overline{\k}} \ll 1/l $ with $ \overline{\k} = (B_{\m\n}B^{\m\n})^{1/2} $ the quadratic mean of the main curvatures. 
With this substitution, we derive
\begin{equation}\label{Wigner-1}
\begin{split}
    W^{\m\n}_\one\big|_T (x,k) = \int_0^1 d\l \int_{P(x)} d^3\ovby \, \bigg\{ 
    - n_\r (\pt_\a\b_\s) (\ovy-x)^\a \lan\opt{T}^{\r\s}(\ovy-i\l\b) \opt{W}^{\m\n} (x,k)\ran_{0,c} \\
     +(\pt_\a\b_\s) B_{\r\t} (\ovy-x)^\t (\ovy-x)^\a \lan\opt{T}^{\r\s}(\ovy-i\l\b) \opt{W}^{\m\n} (x,k)\ran_{0,c} \\
     -\frac{1}{2} n_\r (\pt_\a\b_\s) B_{\t\g} (\ovy-x)^\a (\ovy-x)^\t (\ovy-x)^\g \lan (n\cdot \pt)\opt{T}^{\r\s}(\ovy-i\l\b) \opt{W}^{\m\n} (x,k)\ran_{0,c} \\
     -\frac{1}{2} n_\r n^\a (\pt_\a\b_\s) B_{\t\g} (\ovy-x)^\t (\ovy-x)^\g \lan\opt{T}^{\r\s}(\ovy-i\l\b) \opt{W}^{\m\n} (x,k)\ran_{0,c} 
     + \mc{O}(\pt,\overline{\k}^2)\bigg\}
\end{split}
\end{equation}
The technique to calculate these terms is demonstrated in Ref.~\cite{Zhang:2024mhs}.
Therefore, we only exhibit a simple example here.

Let us focus on the leading term in expression (\ref{Wigner-1}).
Expressing the operators by creators and annihilators and using the Wick theorem, we obtain
{\small
\begin{equation}\label{Wigner-1-expand}
\begin{split}
    W_{\m\n}^\one&\big|_{\pt\b} (x,k) = - n_\r (\pt_\a\b_\s) \int_0^1 d\l \int_{P(x)} d^3\ovby \, \prod_{i=0}^{3}\lb\sum_{a_i}\int\idp{3}{\bp_i}\rb  \\
    & \times (\ovy-x)^{\a} (2\p)^4 \d^{(4)}\lb k-\frac{p_0}{2}-\frac{p_3}{2}\rb \frac{1}{(2E_{\bp_0})(2E_{\bp_3})} e^{i(p_0-p_3)\cdot x}\\
    & \times\Big[ t^{\r\s,\g_1\g_2}(p_1,-p_2) \lan \opt{a}_{\bp_1}^{a_1} \opt{a}^{a_0\dg}_{\bp_0}\ran_0 \lan\opt{a}_{\bp_2}^{a_2\dg}\opt{a}^{a_3}_{\bp_3}\ran_0 \e_{\n}^{a_0*}(p_0)\e_{\g_1}^{a_1}(p_1)\e_{\g_2}^{a_2*}(p_2)\e_{\m}^{a_3}(p_3) e^{-i(p_1-p_2)\cdot y}\\
    & + t^{\r\s,\g_1\g_2}(-p_1,p_2)\lan \opt{a}_{\bp_1}^{a_1\dg} \opt{a}^{a_3}_{\bp_3} \ran_0 \lan \opt{a}_{\bp_2}^{a_2} \opt{a}^{a_0\dg}_{\bp_0} \ran_0 \e_{\n}^{a_0*}(p_0)\e_{\g_1}^{a_1*}(p_1)\e_{\g_2}^{a_2}(p_2)\e_{\m}^{a_3}(p_3) e^{i(p_1-p_2)\cdot y} \Big] \,,
\end{split}
\end{equation}
}
where
\begin{equation}
\begin{split}
    t\,^{\r\s,\g_1\g_2}(p_1,p_2) = \frac{e^{-\l(p_1+p_2)\cdot \b}}{(2E_{\bp_1})(2E_{\bp_2})} \Big[p_1^{\r}p_2^{\s}\w^{\g_1\g_2} -p_1^{\g_2} p_2^{\s}\w^{\r\g_1}\\
     -\frac{1}{2}(p_1\cdot p_2)\w^{\g_1\g_2}\w^{\r\s} +\frac{1}{2} p_1^{\g_2} p_2^{\g_1} \w^{\r\s} -\frac{1}{2}m^2\w^{\g_1\g_2}\w^{\r\s}\Big] \,.
\end{split}
\end{equation}
The first step is to integrate over $ \bp_0 $ and $ \bp_3 $ with related delta functions then to integrate over $ \ovby $ by parts using
\begin{equation}
    \int_{P(x)} d^3\ovby \int \frac{d^3\bp_1}{(2\p)^3} \frac{d^3\bp_2}{(2\p)^3} (\ovy-x)^\a e^{-i(p_1-p_2)\cdot(y-x)} = \int \frac{d^3\bp_1}{(2\p)^3} \frac{d^3\bp_2}{(2\p)^3} (2\p)^3 \d^{(3)}(\bp_1-\bp_2) D^\a(p_1,-p_2) \,,
\end{equation}
where
\begin{equation}
    D^\a(p_1,p_2) = -\frac{i}{2} \D^{\a\m}_\bkn \lb \frac{\pt}{\pt p_1^\m} + \frac{\pt}{\pt p_2^\m} \rb \,.
\end{equation}
Then, we derive from \eq{Wigner-1-expand} by integrating over $ \bp_1 $, $ \bp_2 $ and summing over $ a_i $
\begin{equation}\label{Wigner-Diagram-ptb}
\begin{split}
    W_{\m\n}^\one\big|_{\pt\b} (x,k) =& - (2\p) \d(k^2-m^2) \h(k^0) \, n_\r (\pt_\a\b_\s)  \int_0^1 d\l \, D^{\a}(p_1,-p_2) \frac{2E_\bk}{(2E_{\bp_1})(2E_{\bp_2})}\\
    & \times  G_{\m\g_2}(p_2) \ls t^{\r\s,\g_1\g_2}(p_1,-p_2) + t^{\r\s,\g_2\g_1}(-p_2,p_1) \rs G_{\g_1\n}(-p_1) \bigg|_{p_1=p_2=k} \,,
\end{split}
\end{equation}
where $ G $ stands for the propagator
\[
\begin{split}
    G^{\m\n}(p) &= 2 (n\cdot p) n_B(\b\cdot p) \lb -\w^{\m\n}+\frac{p^\m p^\n}{m^2} \rb \,,\\
    G^{\m\n}(-p) &= 2 (n\cdot p) \ls 1+ n_B(\b\cdot p) \rs \lb -\w^{\m\n}+\frac{p^\m p^\n}{m^2} \rb \,.
\end{split}
\]
$ -n_B(-x) = 1+n_B(x) $ has been used to get the second equation.
To obtain \eq{Wigner-Diagram-ptb}, we use the fact that $ D^\a $ commute with $ \d^{(3)}(\bk - \bp_1/2 - \bp_2/2) $
and omit the off-shell term $ \d'(k^0-E_\bk) $ generated by $ D^\a $ operating on $ \d(k^0 - E_{\bp_1}/2 - E_{\bp_2}/2) $.
In some cases, the off-shell effect would contribute to the spin alignment~\cite{Sheng:2022ssp}.
However, as we consider the free particle, although a local off-shell term appears in the Wigner function,
it is natural that it disappears after we average it over the full hypersurface. It thus does not contribute to $ \H_{rs}(k) $.

Following a similar process, we derive the other terms of $ W^\one_{\m\n}\big|_{T} $ in \eq{Wigner-1}, which are proportional to the curvature tensor,
\begin{equation}\label{Wigner-Diagram-Bptb}
\begin{split}
    W^\one_{\m\n} \big|_{B\pt\b} (x,k) =&~ -(2\p) \d(k^2-m^2) \h(k^0) \, (\pt_\a\b_\s)  \int_0^1 d\l \, \Big[ -B_{\r\t} D^\a D^\t +\frac{1}{2} n_\r n^\a B_{\t\k} \\
    &\times D^\t D^\k +\frac{1}{2} n_\r B_{\t\k} D^\a D^\t D^\k (-i) n\cdot(p_1-p_2) \Big] \frac{2E_\bk}{(2E_{\bp_1})(2E_{\bp_2})} \\
    &\times G_{\m\g_2}(p_2) \ls t^{\r\s,\g_1\g_2}(p_1,-p_2) + t^{\r\s,\g_2\g_1}(-p_2,p_1) \rs G_{\g_1\n}(-p_1) \bigg|_{p_1=p_2=k} \,,
\end{split}
\end{equation}
where $ D^\m $ is short for $ D^\m(p_1, -p_2) $.
With the same assumptions, we evaluate the other terms of $ W^\one_{\m\n} $ generated by the correlator between the spin tensor and the Wigner operator
\begin{align}
    W^{\m\n}_\one\big|_S (x,k) &\equiv \O_{\r\s}(x) \int_0^1 d\l \int_{\S} d\S_\a(y) \lan \opt{S}^{\a\r\s}(y-i\l\b(x)) \opt{W}^{\m\n}(x,k)\ran_{0,c} \\
    &= W^{\m\n}_\one\big|_{\O} (x,k) + W^{\m\n}_\one\big|_{B\O} (x,k) + \mc{O}(\pt,\overline{\k}^2)
\end{align}
with
\begin{equation}\label{Wigner-Diagram-O}
\begin{split}
    W^{\m\n}_\one\big|_{\O} (x,k) =&~ (2\p) \d(k^2-m^2) \h(k^0) \, n_\a \frac{1}{2} \O_{\r\s} \int_0^1 d\l \frac{2E_\bk}{(2E_{\bp_1})(2E_{\bp_2})} \\
    & \times  G_{\m\g_2}(p_2) \ls s^{\a\r\s,\g_1\g_2}(p_1,-p_2) + s^{\a\r\s,\g_2\g_1}(-p_2,p_1) \rs G_{\g_1\n}(-p_1) \bigg|_{p_1=p_2=k}
\end{split}
\end{equation}
and
\begin{equation}\label{Wigner-Diagram-BO}
\begin{split}
     W^{\m\n}_\one\big|_{B\O}& (x,k) = (2\p) \d(k^2-m^2) \h(k^0) \, \frac{1}{2} \O_{\r\s} \int_0^1 d\l  \\
    & \times \Big[ -B_{\a\t} D^\t +\frac{1}{2} n_\a B_{\t\k}D^\t D^\k (-i)n\cdot(p_1-p_2) \Big] \frac{2E_\bk}{(2E_{\bp_1})(2E_{\bp_2})}\\
    & \times G_{\m\g_2}(p_2) \ls s^{\a\r\s,\g_1\g_2}(p_1,-p_2) + s^{\a\r\s,\g_2\g_1}(-p_2,p_1) \rs G_{\g_1\n}(-p_1) \bigg|_{p_1=p_2=k} \,,
\end{split}
\end{equation}
where
\begin{equation}
    s\,^{\a\r\s,\g_1\g_2}(p_1,p_2) = 2i\,\frac{e^{-\l \b\cdot(p_1 +p_2)}}{(2E_{\bp_1})(2E_{\bp_2})} \Big(p_1^{\a}\w^{\r\g_1}\w^{\s\g_2}-p_1^{\r}\w^{\a\g_1}\w^{\s\g_2}\Big) \,.
\end{equation}

In the following, we exhibit the result of $ W^{\m\n}_\one $ by calculating Eqs. (\ref{Wigner-Diagram-ptb}), (\ref{Wigner-Diagram-Bptb}), (\ref{Wigner-Diagram-O}), and (\ref{Wigner-Diagram-BO}).
We first evaluate the derivatives with respect to $ p_1 $ and $ p_2 $, then assign $ p_1 = p_2 = k $ with $ k^\m = (E_\bk,\bk) $, and integrate over $ \l $ at last.
We should emphasize that during the demonstration above, $ p_1 $ and $ p_2 $ should be taken as on-shell, specifically, $ p_1^0 = E_{\bp_1} $ and $ p_2^0 = E_{\bp_2} $.
However, when using a program package to assist in the calculation, \eg~\textsc{FeynCalc}~\cite{Mertig:1990an,Shtabovenko:2016sxi,Shtabovenko:2020gxv}, it is more convenient to regard these two zero components as free variables while modifying $ D^\a(p_1,p_2) $ as
\[
    -\frac{i}{2} \lc \ls\w^{\a\m} - \hat{p}_1^\a n^\m\rs\frac{\pt}{\pt p_1^\m} + \ls\w^{\a\m} - \hat{p}_2^\a n^\m\rs\frac{\pt}{\pt p_2^\m} \rc \,.
\]
We present the result of the projected Wigner function by
\begin{equation}
    W^{\m\n}_{\perp\one}(x,k) =  W^{\m\n}_{\perp\one} \Big|_\pt (x,k) + W^{\m\n}_{\perp\one} \Big|_{B\pt} (x,k) + \mc{O}(\pt,\overline{\k}^2) \,,
\end{equation}
where
\begin{equation}\label{result-Wigner-pt}
\begin{split}
    W^{\m\n}_{\perp\one} \Big|_\pt (x,k) =& -i (2\p) \d(k^2-m^2) \h(k^0) n_B (1+n_B) \D^{\m\r}_\bkk\D^{\n\s}_\bkk \\
    & \times \ls \varpi_{\r\s}(x)
    - \tensor{\Xi}{^\a_{[\r}} \lb \x_{\s]\a}(x) + \d\O_{\s]\a}(x)\rb\rs
\end{split}
\end{equation}
with $ \Xi_{\m\n} = \w_{\m\n} - \hat{k}_\m n_\n $, $ \varpi_{\m\n} = \pt_{[\n}\b_{\m]} $ the thermal vorticity, $ \x_{\m\n} =\pt_{(\m}\b_{\n)} $ the thermal shear tensor, $ \d\O_{\m\n} = \O_{\m\n} - \varpi_{\m\n} $ the net spin potential, and
\begin{equation}\label{result-Wigner-Bpt}
\begin{split}
    W^{\m\n}_{\perp\one} \Big|_{B\pt} (x,k) =&~ (2\p) \d(k^2-m^2) \h(k^0) n_B(1+n_B) \frac{B_{\a\b}}{2(n\cdot k)} \bigg[ \x_{\r\s}\lb\w^{\r\s}+\frac{k^\r k^\s}{2m^2}\rb \\
    & \times \Xi^{\a\m} \Xi^{\b\n} + \lb \d\O_{\r\s} -\x_{\r\s} \rb \Xi^{\a\r} \Xi^{\b(\m} \D^{\n)\s}_{\bkk} + \lb \D^{\m\n}_{\bkk} \cdots\rb \bigg] \,.
\end{split}
\end{equation}
For simplicity, we have not displayed the terms proportional to $ \D^\bkk_{\m\n} $ in \eq{result-Wigner-Bpt}.
In the derivative expansion, these terms contribute to the scalar distribution at $ \mathcal{O}(\pt,\overline{\k})$.
According to \eq{extract-tensor}, they would induce tensor polarization only at subleading order.

Equation (\ref{result-Wigner-pt}) has been obtained for a different scenario in Ref.~\cite{Zhang:2024mhs}.
Since $ W^{\m\n}_{\one} |_\pt $ is antisymmetric concerning $ \m $ and $ \n $, it would only contribute to vectorial spin polarization, not scalar distribution, nor tensor polarization.
$ W^{\m\n}_\one |_{B\pt} $ is symmetric concerning the Lorentz indices; therefore, it would contribute to scalar and tensor distribution functions.
Besides, up to the leading order, the effect of curvature enters only if $ \x \neq 0 $ or $ \O \neq \varpi $.
We present a demonstration by turning off the thermal shear and the net spin potential and expanding the density operator up to $ \mc{O}(\pt) $
\[
    \opt{\r}_\LE \approx \frac{1}{Z} \exp \lc -\b(x) \cdot \opt{P} +\frac{1}{2} \varpi_{\r\s}(x) \opt{J}^{\r\s}(x)\rc \,,
\]
where $ \opt{J}^{\r\s}(x) $ is the total angular momentum with respect to point $ x $, which is defined by
\[
    \opt{J}^{\r\s}(x) \equiv \int_\S d\S_\m(y) \ls (y-x)^\r \opt{T}^{\m\s}(y) -(y-x)^\s \opt{T}^{\m\r}(y) +\opt{S}^{\m\r\s}(y) \rs \,.
\]
Since the total angular momentum is defined by the integral of a conserved current, it does not depend on the geometry of the hypersurface.
Therefore, the effect of curvature disappears if the fluid reaches the global equilibrium.

By combining Eqs. (\ref{result-Wigner-pt}) and (\ref{result-Wigner-Bpt}) and applying Eqs. (\ref{extract-vector}) and (\ref{extract-tensor}), we extract the spin polarization vector
\begin{equation}
    \mathbb{S}^\m (x,k) = -\frac{1+n_B}{3m} \e^{\m\n\r\s} k_\n \ls \varpi_{\r\s}
    - \Xi^\a{}_{[\r} \lb \x_{\s]\a} + \d\O_{\s]\a}\rb\rs +\mc{O}(\pt,\overline{\k}^2)
\end{equation}
and the spin polarization tensor
\begin{equation}\label{result-T}
\begin{split}
     \mathbb{T}^{\m\n} (x,k) =&~ \frac{1+n_B}{6(n\cdot k)} B_{\a\b}\bigg\{-\x_{\r\s}\lb\w^{\r\s}+\frac{k^\r k^\s}{2m^2}\rb \ls \Xi^{\a\m} \Xi^{\b\n} -\frac{1}{3}\D_\bkk^{\m\n}\lb\w^{\a\b}+\hat{k}^\a \hat{k}^\b\rb \rs \\
     & + \x_{\r\s} \ls \Xi^{\a\r} \Xi^{\b(\m} \D^{\n)\s}_{\bkk} -\frac{1}{3} \D_\bkk^{\m\n} \Xi^{\a\r}\Xi^{\b\s} \rs -\d\O_{\r\s} \Xi^{\a\r} \Xi^{\b(\m} \D^{\n)\s}_{\bkk} \bigg\} +\mc{O}(\pt,\overline{\k}^2) \,.
\end{split}
\end{equation}
These expressions describe the polarizations generated by first-order hydrodynamic fields in $\pt$.
Equation (\ref{result-T}) presents the tensor spin polarization induced by first-order hydrodynamic fields with a curved hypersurface.
The curvature appears at leading order.
We treat curvature as an independent order, since it is a general feature in our analysis.
In practice, the freeze-out hypersurface is often chosen as a constant-$ T $ surface.
For this case, the curvature is an $ \mc{O}(\pt) $ term.
Therefore, tensor polarization in \eq{result-T} is of the order of $\mc{O}(\pt^2) $, same as the ones in Ref.~\cite{Zhang:2024mhs,Yang:2024fkn}.
However, the curvature-related term is new and absent in prior literature.
This term originates from statistical effects and depends on the definition of the local equilibrium on the freeze-out hypersurface.

\section{Freeze-Out in Bjorken Expansion}
\label{sec:Bjorken}
In the following, we estimate \eq{result-T} in the flow with Bjorken expansion~\cite{Bjorken:1982qr}.
Since in the Bjorken expansion, the fluid preserves the longitudinal boost invariance, we should expect the freeze-out to take place on a hypersurface with a constant ``proper time" $ \t = \sqrt{t^2-z^2} $, which is described by 
\[
    2F(x) = g^\parallel_{\m\n} x^\m x^\n - \t_f^2 = 0 \,,
\]
where $ g^\parallel_{\m\n} = \diag\lb1,0,0,-1\rb $.
Based on $ F(x) $, we have the normal vector
\begin{equation}
    n^\m = x_\parallel^\m / \t_f
\end{equation}
with $ x^\parallel_\m = g^\parallel_{\m\n}x^\n $
and the curvature tensor
\begin{equation}\label{Bjorken-curvature}
    B^{\m\n} = (n^\m n^\n - g_\parallel^{\m\n})/\t_f \,.
\end{equation}
As hinted by Bjorken expansion, the fluid velocity is $ u_\m = x^\parallel_\m/\t $ and the gradient of thermal velocity reads
\[
    \pt_\m\b_\n = \frac{1}{T} \lb \pt_\m u_\n - \frac{u_\m u_\n}{T}\frac{dT}{d\t} \rb
\]
with $ \pt_\m u_\n = \lb g^\parallel_{\m\n} - u_\m u_\n \rb / \t $ and the cooling rate $ dT/\lb Td\t \rb = -c_s^2 / \t $, where $ c_s $ is the speed of sound.
The cooling rate is obtained in Ref.~\cite{Bjorken:1982qr} by using the ideal hydrodynamic equation and assuming the zero chemical potential, which is applicable for higher collision energy in RHIC.
Therefore, the thermal vorticity is zero, and the thermal shear reads
\begin{equation}\label{Bjorken-shear}
    \x^{\m\n} = \frac{1}{T\t} \ls g_\parallel^{\m\n} - (1 - c_s^2) u^\m u^\n \rs \,
\end{equation}

Plugging Eqs. (\ref{Bjorken-curvature}) and (\ref{Bjorken-shear}) into \eq{result-T} and turning the net spin potential to zero, one can write out the tensor polarization in Bjorken expansion.
To estimate the result, we take another primary assumption that the particle produced is static inside the fluid cell, namely, $ k_\m = m u_\m $.
Then, we obtain
\begin{equation}\label{Bjorken-T}
    \mathbb{T}^{\m\n} (x,k) = -\frac{(1+n_B) c_s^2}{4m T_f \t_f} B^{\lan\m\n\ran} \,,
\end{equation}
where $ T_f $ is the temperature when the vector mesons freeze out.
$ B_{\lan\m\n\ran} $ implies that in this typical situation, the tensor polarization comes from the anisotropy of the curvature of the freeze-out hypersurface.
By taking $ T_f $ to be $ 150 \text{ MeV} $, $ \t_f $ to be $ 5 \text{ fm} $, and $ m $ to be $ \phi $ meson's mass $ 1019 \text{ MeV} $,
we estimate the magnitude of the spin alignment 
\begin{equation}\label{Bjorken-H}
    \d\H_{yy} (x,k) = -\frac{(1+n_B) c_s^2}{12 m T_f \t_f^2} \sim -10^{-4} \,.
\end{equation}
From the symmetry between $ x $ and $ y $, we can also conclude that $ \d \H_{xx} < 0 $, and thus, $ \d\H_{zz} > 0 $. 
Therefore, Eq.~(\ref{Bjorken-T}) seems to imply that in this situation, the longitudinal polarization state of a vector meson tends to align with the direction where the curvature is greater than the average, namely, the direction where the hypersurface curves more toward the normal vector, in this case, $ z $ direction.
This leads to a smaller probability of longitudinal polarization in the $ y $ direction, namely, $ \d\H_{yy} < 0 $.

It is useful to compare \eq{Bjorken-H} with existing measurements. Although the $ \d\H_{yy} $ signal of the $\phi$ meson can reach \(\sim 0.03\) at lower beam energies~\cite{STAR:2022fan}, the observed spin alignment in collisions closer to the Bjorken assumptions, namely higher-energy collisions, is much smaller, typically at the \(\sim 10^{-3}\) level. In this sense, the estimate in \eq{Bjorken-H} should be viewed only as a small negative baseline contribution to the spin-alignment signal in high-energy collisions.
\section{Freeze-Out in Gubser Expansion}
\label{sec:Gubser}
Gubser flow is an analytic solution to the conformal invariant Navier-Stokes equation~\cite{Gubser:2010ze}.
It possesses longitudinal boost invariance and rotational symmetry about the $ z $ axis (corresponding to the central collisions).
However, compared to the Bjorken flow, it lacks transverse translational symmetry,
which accommodates the finite transverse size in realistic collisions and thus qualitatively depicts the radial flow.

Gubser flow contains a free parameter $ q $, whose inverse characterizes the transverse size of the system.
In the following, we adopt $ 1/q = 4.3 \text{ fm} $, which is obtained by estimating the gold-gold central collisions at RHIC energy~\cite{Gubser:2008pc,Gubser:2010ze}.
To cooperate with these symmetry properties, we adopt a reference frame $ (\t,\w,\r,\f) $ with $ \t $ the ``proper time" defined in Sec.~\ref{sec:Bjorken},
$ \w $ the space-time rapidity, $ \r = \sqrt{x^2+y^2} $ the transverse radius, and $ \f $ the azimuthal angle.
We follow the result of inviscid fluid in Ref.~\cite{Gubser:2010ze}
\begin{align}
\label{Gubser-temperature}
    T &= \frac{\Tilde{T}_0}{\t (1+g^2)^{1/3}} \,, \\
\label{Gubser-velocity}
    u^\m &= \frac{1+q^2\t^2 +q^2\r^2}{2q\t\sqrt{1+g^2}} \frac{\pt x^\m}{\pt\t} + \frac{q\r}{\sqrt{1+g^2}}\frac{\pt x^\m}{\pt \r} \,,
\end{align}
where 
\[
    g = \frac{1-q^2\t^2+q^2\r^2}{2q\t} \,,
\]
and $ {\pt x^\m}/{\pt\t} $ and $ {\pt x^\m}/{\pt \r} $ stand for unit vectors in $ \t $ and $ \r $ directions, namely,
\[
    \frac{\pt x^\m}{\pt\t} = \frac{1}{\t} x^\m_\parallel \,,\quad \frac{\pt x^\m}{\pt \r} = \frac{1}{\r} x_\perp^\m
\]
with $ x_\perp^\m = g^{\m\n}_\perp x_\n $ and $ g_\perp^{\m\n} = \diag(0,-1,-1,0) $.
According to a quasirealistic estimation in Ref.~\cite{Gubser:2010ze}, for gold-gold collisions at RHIC energy, $ \Tilde{T}_0 = 2.99 $.
Letting $ q \rightarrow 0 $ properly (keeping $ q\Tilde{T}_0 $ a constant), one would recover the Bjorken flow.

If we assume the freeze-out temperature to be $ 150 \text{ MeV} $, given by \eq{Gubser-temperature}, $ q\t \sim 0.91 $ in the paraxial region ($ q\r \ll 1 $) of the freeze-out hypersurface.
Therefore, to simplify the calculation, we focus on the paraxial region of the hypersurface, and thus,
we assume $ \d\Tilde{\t} = q\t-1 $ and $ \d\Tilde{\r} = q\r $ as small quantities $ \d \sim 0.1 $ and continue our calculation by performing these expansions.
For example, for temperature and fluid velocity, we have
\begin{align*}
    T &= \Tilde{T}_0 q \ls 1 -\d \Tilde{\t} +\frac{2}{3} \d \Tilde{\t}^2 -\frac{1}{3} \d \Tilde{\t} \lb \d \Tilde{\t}^2 -\d \Tilde{\r}^2\rb \rs + \mc{O}(\d^4)\,, \\
    u^\m &= \lb 1+\frac{1}{2} \d \Tilde{\r}^2 \rb \frac{\pt x^\m}{\pt\t} + \d\Tilde{\r} \frac{\pt x^\m}{\pt \r} + \mc{O}(\d^4) \,.
\end{align*}
Since we will use the second-order spatial derivative of the temperature field, we have to expand it up to $ \d^3 $.

With the equation of hypersurface, $ \t ^3 (1+g^2) = \text{const.} $, we obtain
\[
    n^\m = \frac{\pt x^\m}{\pt \t} -\frac{2}{3} \d\Tilde{\t} \d \Tilde{\r} \frac{\pt x^\m}{\pt \r} +\mc{O}(\d^3)\,,
\]
and thus we obtain the curvature tensor
\begin{equation}\label{Gubser-curvature}
    B^{\m\n} \equiv -\D^{\m\a}_\bkn \frac{\pt n^\n}{\pt x^\a} = \frac{1}{\t}\lb n^\m n^\n -g^{\m\n}_\parallel\rb -\frac{2}{3} q \d\Tilde{\t} g_\perp^{\m\n} +\mc{O}(\d^2)\,.
\end{equation}
As for the thermal shear tensor, we first calculate the cooling rate
\[
    \frac{(n\cdot\pt)T}{T} = -q\lb 1-\frac{1}{3}\d\Tilde{\t}\rb +\mc{O}(\d^2)
\]
and the kinetic shear $ \pt_{(\m}u_{\n)} $, then we have
\begin{align}
    \x^{\m\n} &= \frac{1}{T}\lb \pt^{(\m}u^{\n)} -u^{(\m} n^{\n)} \frac{(n\cdot\pt)T}{T} \rb  \non
\label{Gubser-shear}
    &= \frac{1}{\Tilde{T}_0} \ls \w^{\m\n} + \d\Tilde{\t}\lb g^{\m\n}_\perp +\frac{2}{3}n^\m n^\n \rb\rs +\mc{O}(\d^2) \,.
\end{align}
As in the previous section, we assume $ k_\m = m u_\m $. Plugging Eqs. (\ref{Gubser-curvature}) and (\ref{Gubser-shear}) into \eq{result-T}, we have
\begin{equation}\label{Gubser-T}
    \mathbb{T}^{\m\n}(x,k) = -\frac{1+n_B}{6 m \Tilde{T}_0} \lb \frac{7}{2} + \d\Tilde{\t} \rb B^{\lan \m\n \ran} (x) \,,
\end{equation}
where $ B^{\lan \m\n \ran} $ implies that in the Gubser flow and with all those assumptions, an anisotropic curvature would induce tensor polarization.
Using the definition of $ \e_y(k) $, we obtain the spin alignment measured in the $ y $ direction
\begin{equation}\label{Gubser-spa}
    \d\H_{yy}(x,k) = -\frac{(1+n_B)q}{6 m \Tilde{T}_0} \lb \frac{7}{6} -\frac{29}{18}\d\Tilde{\t} \rb \sim - 3 \times 10^{-3} \,.
\end{equation}
Therefore, similar to the Bjorken flow, in this situation, the spin $ \pm 1 $ state of a vector meson tends to align in the direction with smaller curvature than the average.
Thus, the probability of longitudinal polarization in $ x $ and $ y $ directions is lower than $ z $.

Compared with the Bjorken estimate~\eqref{Bjorken-H}, the Gubser estimate~\eqref{Gubser-spa} is enhanced by the finite transverse size and reaches the $10^{-3}$ level. Nevertheless, it remains negative and is more appropriately regarded as a benchmark contribution rather than a complete description of the data.

From \eq{Gubser-spa}, one can also observe a property by choosing a system with a different size $ L $,
such as a lead-lead collision or a slightly noncentral gold-gold collision.
Since $ 1/q $ depicts the typical transverse length of the system, one should expect $ q \sim L^{-1} $.
And according to Ref.~\cite{Gubser:2010ze}, $ \Tilde{T}_0 \sim (dN/d\w)^{1/3} $. Since $ dN/d\w \sim L^{3} $, we have $ \Tilde{T}_0 \sim L $.
From the leading term of Eq.~\eqref{Gubser-spa}, one finds \( \d\H_{yy} \sim L^{-2} \), implying a stronger curvature effect for a smaller system. Based on this scaling, we can make a rough estimate for central O-O collisions. Taking \(L_{\text{Au}} : L_{\text{O}} \sim 2.5\), we obtain
\begin{equation}
    \d\H_{yy}^{(\mathrm{O})}\sim -10^{-2} \,.
\end{equation}
This estimate is negative and sizable, exceeding in magnitude the spin-alignment signal observed for the $\phi$ meson in higher-energy Au-Au collisions, where the Gubser picture is expected to be more relevant. If other contributions to spin alignment, such as those from the $\phi$ mean field, do not gain a comparable enhancement in high-energy O-O collisions, the curvature-induced contribution may become more prominent there. In this sense, spin-alignment measurements in high-energy O-O collisions may provide a relatively clean probe of the polarization induced by the curvature of the freeze-out hypersurface.

\section{Conclusion and Discussion}
\label{sec:conclusion}
In this work, we formulate a covariant description of spin polarization for massive vector bosons using components of the Wigner function, $ \mathbb{S}^\m(x,k) $ and $ \mathbb{T}^{\m\n}(x,k) $, defined in \eqs{extract-scalar}{extract-tensor}.
For instance, the spin polarization tensor is given by the traceless and symmetric part of the projected Wigner function
\begin{equation}
    \mathbb{T}^{\m\n}(x,k) = \frac{W^{\lan\m\n\ran}_{\perp}(x,k)}{W^{\r}_{\perp\r}(x,k) }  \,.
\end{equation}
By evaluating its average over the hypersurface via the Cooper-Frye formula (\ref{cooper-frye-tensor}),
we obtain the expectation value of the spin polarization tensor (\ref{def-spin_tensor_operator}).
Our results show that the curved freeze-out hypersurface significantly affects the spin alignment of massive vector bosons in relativistic fluids.
We identify the curvature tensor $ B_{\m\n} $ defined in \eq{def-curv} as a key contributor to the tensor polarization.
Specifically, curvature-induced contributions arise from terms proportional to first-order hydrodynamic fields, such as the thermal shear $ \x_{\r\s} $ and the net spin potential $ \d\O_{\r\s} $
\begin{multline}
    \mathbb{T}^{\m\n} (x,k) = \frac{1+n_B}{6(n\cdot k)} B_{\a\b}\bigg[ -\x_{\r\s}\lb\w^{\r\s}+\frac{k^\r k^\s}{2m^2}\rb \Xi^{\a\langle\m} \Xi^{\b\n\rangle} \\
    + \x_{\r\s} \Xi^{\a\r} \Xi^{\b\langle\m} \D^{\n\rangle\s}_{\bkk} -\d\O_{\r\s} \Xi^{\a\r} \Xi^{\b(\m} \D^{\n)\s}_{\bkk} \bigg] \,.
\end{multline}
We estimate the tensor polarization for both Bjorken and Gubser flows, revealing that anisotropic curvature induces tensor polarization.
Notably, the longitudinal polarization state prefers to align with directions of greater curvature.
Consequently, the spin alignment along the $ y $ direction, $ \d \H_{yy} $, is negative and with a magnitude of $ 10^{-4} - 10^{-3} $.
Further analysis of the Gubser flow suggests that curvature effects on spin alignment are more significant in systems with smaller transverse sizes.

We stress that the above discussion applies specifically to the spacelike freeze-out hypersurface.
For implementation in numerical simulations, future research must address the timelike part of the freeze-out hypersurface.
Additionally, the above result depends on our choice of pseudo-gauge. In particular, for the tensor polarization, we emphasize that the results differ from the Belinfante pseudo-gauge even if we neutralize the net spin potential, namely, setting the spin potential to be the thermal vorticity. This occurs because the canonical spin tensor for the vector boson is not fully antisymmetric~\cite{Becattini:2018duy}. These issues will be considered in future studies.

\section*{Acknowledgments}
 Z.-H.Z. and X.-G.H. are supported by the Natural Science Foundation of Shanghai (Grant No. 23JC1400200), the National Natural Science Foundation of China (Grant No. 12225502 and No. 12147101), and the National Key Research and Development Program of China (Grant No. 2022YFA1604900).

\appendix
\section{Spin polarization tensor}
\label{appx:covariant-tensor}
For a single particle state in relativistic quantum mechanics, the spin polarization vector $ \mS^\m (k) $ defined by \eq{covariant-polar-vector} is the expectation value of the vectorial spin operator $ \opt{\mS}^\m $ defined by \eq{def-spin_vector_operator}, which has been proved in Ref. \cite{Becattini:2020sww}.
In this appendix, we follow Ref. \cite{Becattini:2020sww} to demonstrate that for a single massive particle state, $ \mT^{\m\n} (k) $ defined by \eq{covariant-polar-tensor} is the expectation value of the tensor spin operator $ \opt{\mT}^{\m\n} $ defined by \eq{def-spin_tensor_operator}.

For a single massive particle, the eigenspace of the momentum operator $ \opt{P}_\m $ is formed by $ \ket{k} $ that satisfy $ \opt{P}_\m \ket{k} = k_\m \ket{k} $,
where $ k_\m $ is on the mass shell, $ k^0 = E_\bk $.
Noticing that $ [\opt{\mS}_\m , \opt{P}_\n] = 0 $, we can restrict $ \opt{\mS}_\m $ to the eigenspace of $ \opt{P}_\m $ labeled by $ k_\m $.
As $ \opt{\mS} \cdot \opt{P} = 0 $, for all the $ \ket{k} $ in the eigenspace, we have $ \opt{\mS}\cdot k \ket{k} = 0 $.
Hence, we can conclude that
\[
    \opt{\mS}(k)\cdot k = 0 \,,
\]
where $ \opt{\mS}_\m(k) $ represents the restriction of $ \opt{\mS}_\m $ mentioned earlier.
Therefore, we can decompose $ \opt{\mS}_\m (k) $ by polarization vectors $ \e^\m_i(k) $
\begin{equation}
    \opt{\mS}^\m (k) = \sum_{i=1}^{3} \opt{\mS}^i(k) \, \e_i^\m(k) \,.
\end{equation}
Similarly, $ \opt{\mT}^{\m\n} $ defined by \eq{def-spin_tensor_operator} turns into
\[
    \opt{\mT}^{\m\n}(k) = \opt{\mS}^{(\m}(k) \opt{\mS}^{\n)}(k) + \frac{2}{3} \D^{\m\n}_\bkk
\]
after being restricted in the eigenspace, which can also be decomposed by polarization vectors
\begin{equation}
    \opt{\mT}^{\m\n}(k) = \sum_{ij} \lb \opt{\mS}^{(i}(k) \opt{\mS}^{j)}(k) -\frac{2}{3} \d^{ij} \rb \e_i^\m(k) \e_j^\n(k) \,.
\end{equation}

Since $ [\opt{P}^\m , \opt{\mS}^i(k)] = 0 $, we can choose to diagonalize one of $ \opt{\mS}^i (k) $ in the eigenspace of $ \opt{P}_\m $.
Instead, we choose the adjoint basis, namely,
\[
    \opt{P}_\m \ket{k,i} = k_\m \ket{k,i} \,, \quad \opt{\mS}^i(k) \ket{k,i}= 0 \,, \qquad \text{with } i = 1,2,3 \,,
\]
where we choose to characterize the eigenspace by the longitudinal states of $ \opt{\mS}^i(k) $.
Since $ \opt{\mS}^i(k) $ are the generators of the little group of massive particles, namely, the $ SU(2) $ group, we have
\begin{equation}
    \bra{k,r} \opt{\mS}^i(k) \ket{k,s} = (s^{i})_{rs} \,,
\end{equation}
where $ s^{i} $ are defined by $ (s^{i})_{jk} = -i \e^{ijk} $ as we take the adjoint representation here.
Similarly, we obtain
\begin{equation}
    \bra{k,r} \big[\opt{\mS}^{(i}(k) \opt{\mS}^{j)}(k) -\frac{2}{3} \d^{ij} \big]\ket{k,s} = (s^{(i}s^{j)})_{rs} - \frac{2}{3}\d^{ij}\d_{rs} = (\S^{ij})_{rs}
\end{equation}
for the tensor spin operator.

Let $ \opt{\r} $ be the density operator of the single particle in the Hilbert space, and its restriction to the eigenspace of $ \opt{P}_\m $ labeled by $ k_\m $ be $ \opt{\H}(k) $, which is precisely the spin density operator of a particle with momentum $ \bk $.
Evaluating the expectation value of $ \opt{\mT}^{\m\n}(k) $, we obtain
\begin{align*}
    \Tr{\opt{\H}(k) \opt{\mT}^{\m\n}(k)} =&~ \sum_{r,s} \bra{k,r} \opt{\H}(k) \ket{k,s} \bra{k,s} \opt{\mT}^{\m\n}(k) \ket{k,r} \\
    =&~ \sum_{r,s,i,j} \H_{rs}(k) (\S^{ij})_{sr} \e_i^\m(k) \e_j^\n(k) = \sum_{i,j} \fT^{ij}(k) \e_i^\m(k) \e_j^\n(k)
\end{align*}
where we utilize \eq{tensor-polar} in the last equation.
Comparing this with \eq{covariant-polar-tensor}, we conclude that for a single particle state, $ \mT^{\m\n} (k) $ defined by \eq{covariant-polar-tensor} is the expectation value of the tensor spin operator $ \opt{\mT}^{\m\n} $ defined by \eq{def-spin_tensor_operator}.

\section{Geometric interpretation of the curvature tensor}
\label{appx:curv}
In this appendix, we clarify the geometric meaning of the curvature tensor $B_{\mu\nu}$ introduced in \eq{def-curv}.
More precisely, $B_{\mu\nu}$ is the extrinsic curvature tensor of the hypersurface $\Sigma$. Its contraction with a unit tangent vector gives the normal curvature of $\Sigma$ in that direction.
Its traceless part characterizes the curvature anisotropy relevant for the tensor polarization in \eq{Bjorken-T} and \eq{Gubser-T}.

To see these, let us consider a point $x\in\Sigma$ and a unit tangent vector $\zeta^\mu$ at $x$, satisfying
\(\zeta\cdot n=0\) and \(\zeta^2=-1\).
To characterize how the hypersurface bends along the direction $\zeta^\mu$, we consider the two-dimensional plane spanned by $\zeta^\mu$ and the hypersurface normal $n^\mu$.
The intersection of this plane with $\Sigma$ defines a local curve through $x$, known as the normal section of $\Sigma$ associated with $\zeta^\mu$.
Since this is an ordinary plane curve, its curvature can be defined in the usual way.
Let $X^\mu(s)$ denote this curve, parametrized by its arc-length $s$, and let
\[
    t^\mu(s)\equiv \frac{dX^\mu}{ds}
\]
be its unit tangent vector. At the point $x$, one has \(X^\mu(0)=x^\mu \) and
\( t^\mu(0)=\zeta^\mu \).
Because \(t^\mu t_\mu\) is constant, one has \( t_\mu dt^\mu / ds = 0 \),
so $dt^\mu/ds$ is orthogonal to $t^\mu$. For the normal section, the relevant normal
direction within the plane is precisely the hypersurface normal $n^\mu$.
Thus, the turning rate of the tangent direction $ dt^\mu/ds $ is parallel to $ n^\mu $.
The curvature of the normal section at $x$ is the rate at which the tangent
direction turns toward the normal direction. Therefore, the normal curvature
of the normal section along the tangent direction $\zeta^\mu$ is given by
\begin{equation}\label{curvature-definition}
    \kappa(\zeta)=n_\mu \frac{dt^\mu}{ds}\bigg|_{s=0} \,.
\end{equation}

Since the curve lies on the hypersurface, its tangent remains orthogonal to the hypersurface normal
\[
    n_\mu(X(s))\, t^\mu(s)=0 \,.
\]
Differentiating this relation with respect to $s$ and plugging in \eq{curvature-definition},
we obtain
\[
    \k(\z) = - \z^\m \frac{d}{ds}n_\m(X(s))\bigg|_{s=0} \,.
\]
Using the chain rule, \( d n_\m(X(s)) /ds = t^\alpha \partial_\alpha n_\mu\), we arrive at
\begin{equation}
    \k(\z) = - \z^\a \z^\m \, \pt_\a n_\m \,.
\end{equation}
Finally, since $\zeta^\mu$ is tangent to the hypersurface, one has
$\zeta^\mu \Delta^{(n)}_{\mu\alpha}=\zeta_\alpha$.
And thus, with the definition~\eqref{def-curv}, we obtain
\begin{equation}
    \kappa(\zeta)=B_{\mu\nu}\zeta^\mu \zeta^\nu \,.
\end{equation}
Therefore, the quadratic form defined by $B_{\mu\nu}$ gives the normal curvature
of the hypersurface along the tangent direction $\zeta^\mu$.

Since $B_{\mu\nu}$ is symmetric and acts on the tangent space of the hypersurface,
its eigenvectors define the principal directions, and the corresponding eigenvalues
are the principal curvatures.
In particular,
\[
    B_{\m\n} = \sum_{i=1}^3 \k(\z^{(i)}) \z^{(i)}_\m \z^{(i)}_\n \,,
\]
where $\{\zeta^{(i)\mu}\}$ is an orthonormal basis of tangent space formed by principal directions. The traceless part
\begin{equation}\label{curv-traceless}
    B_{\m\n} - \frac{1}{3} \D^{(n)}_{\m\n} B^{\a}{}_{\a}
\end{equation}
therefore removes the isotropic average curvature and isolates the directional dependence of the local bending, namely the local curvature anisotropy.
In \eq{Bjorken-T}, $B_{\lan\mu\nu\ran}$ is precisely the traceless tensor in \eqref{curv-traceless}, since in the derivation of \eq{Bjorken-T} we have assumed $k^\mu/m=u^\mu$, and for the Bjorken flow one has $u^\mu=n^\mu$.
For the Gubser flow, by contrast, $ u^\m = n^\m +\mc{O}(\d) $. Accordingly, the tensor $B_{\lan\mu\nu\ran}$ appearing in \eq{Gubser-T} reduces to \eqref{curv-traceless} in the limit $\delta\to0$.

\bibliography{biblio}

\end{document}